%% file: main.tex
\pdfoutput=1

\documentclass[sigconf,nonacm]{acmart}

\usepackage{color}
\definecolor{grau}{RGB}{136,136,136}
\usepackage{pgf-pie} 
\usepackage{longtable}
\usepackage{supertabular}
\usepackage{tikz}
\usepackage{pgfplots}
\pgfplotsset{compat=1.7}
\usepackage{xcolor}
\usepackage{pgfplotstable}
\usepackage{colortbl}
\usepackage{amsmath}
\usepackage{placeins}

\usepackage{ifthen}
\newboolean{long} 

\setboolean{long}{true}  

\newcommand{\longVersionURL}{\url{https://arxiv.org/exampleURL}}

\AtBeginDocument{%
  \providecommand\BibTeX{{%
    \normalfont B\kern-0.5em{\scshape i\kern-0.25em b}\kern-0.8em\TeX}}}

\setcopyright{acmcopyright}
\copyrightyear{2018}
\acmYear{2018}
\acmDOI{XXXXXXX.XXXXXXX}

\acmConference[Conference acronym 'XX]{Make sure to enter the correct
  conference title from your rights confirmation emai}{June 03--05,
  2018}{Woodstock, NY}
\acmBooktitle{Woodstock '18: ACM Symposium on Neural Gaze Detection,
 June 03--05, 2018, Woodstock, NY} 
\acmPrice{15.00}
\acmISBN{978-1-4503-XXXX-X/18/06}

\begin{document}

\ifthenelse{\boolean{long}}{
\title[$crypto_{lib}$ index]{$crypto_{lib}$: Comparing and selecting cryptography libraries \\(long version of EICC 2022 publication)}
\setcopyright{none}
}{
\title[$crypto_{li b}$ index]{$crypto_{lib}$: Comparing and selecting cryptography libraries}
}
\author{Jan Wohlwender}
\orcid{0000-0001-8647-6322}
\authornote{Both authors contributed equally to this research.}
\affiliation{%
  \institution{Hochschule Darmstadt}
  \streetaddress{Schöfferstraße 3}
  \city{Darmstadt}
  \country{Germany}
  \postcode{64295}
}

\author{Rolf Huesmann}
\orcid{0000-0003-0798-2919}
\authornotemark[1]
\affiliation{%
  \institution{Hochschule Darmstadt}
  \streetaddress{Schöfferstraße 3}
  \city{Darmstadt}
  \country{Germany}
  \postcode{64295}
}

\author{Andreas Heinemann}
\orcid{0000-0003-0240-399X}
\affiliation{%
  \institution{Hochschule Darmstadt}
  \streetaddress{Schöfferstraße 3}
  \city{Darmstadt}
  \country{Germany}
  \postcode{64295}
}

\author{Alexander Wiesmaier}
\orcid{0000-0002-1144-549X}
\affiliation{%
  \institution{Hochschule Darmstadt}
  \streetaddress{Schöfferstraße 3}
  \city{Darmstadt}
  \country{Germany}
  \postcode{64295}
}

\renewcommand{\shortauthors}{Wohlwender and Huesmann, et al.}

\begin{abstract} 
Selecting a library out of numerous candidates can be a laborious and resource-intensive task. We present the $crypto_{lib}$ index, a tool for decision-makers to choose the best fitting cryptography library for a given context.
To define our index, $15$ library attributes were synthesized from findings based on a literature review and interviews with decision-makers.
These attributes were afterwards validated and weighted via an online survey. 
In order to create the index value for a given library, the individual attributes are assessed using given evaluation criteria associated with the respective attribute.
As a proof of concept and to give a practical usage example, the derivation of the $crypto_{lib}$ values for the libraries Bouncy Castle and Tink are shown in detail.
 Overall, by tailoring the weighting of the $crypto_{lib}$ attributes to their current use case, decision-makers are enabled to systematically select a cryptography library fitting best to their software project at hand in a guided, repeatable and reliable way. 
\end{abstract}



\maketitle

\section{Introduction and outline}
In computer science, it is common to outsource often used functionalities to libraries.
These are then included in software projects if required.
A particular advantage of cryptography libraries is the professional implementation and maintenance of cryptographic functionalities by cryptography experienced developers.
Non-cryp\-to\-graphy experienced developers can use these libraries without having to deal with the rather complicated mathematical fundamentals and other obstacles of cryptography. 
Currently, there are numerous cryptography libraries\footnote{\url{https://en.wikipedia.org/wiki/Comparison\_of\_cryptography\_libraries} Retrieved 03~Jan 2022} to choose from.
This work defines the $crypto_{lib}$ index to compare and select cryptography libraries based on a guided evaluation of weighted attributes.
This index can be used efficiently by decision-makers to select a cryptography library for their software project.

We first present related work in Section \ref{ch:verwandte} that also describes contributions to library selection.
Then, attributes and associated evaluation criteria that allow a rating to be made are identified.
The attributes are first compiled in Section~\ref{ch:attribute:literatur} based on a literature research and afterwards deepened in Section~\ref{ch:attribute:interview} based on interviews with decision-makers of software projects.
In Section~\ref{ch:attribute:synthese} these attributes are then synthesized into a manageable set.
Section~\ref{ch:attribute:validierung} describes the conducted validation of the synthesized set through an online survey.
The $crypto_{lib}$ index is introduced in Section~\ref{ch:definition} and described in detail in Section~\ref{ch:index:beschreibung}. 
The calculated index values for the libraries Bouncy Castle and Tink are presented and discussed
in Section~\ref{ch:index:beispiel}.
We discuss our results in Section~\ref{ch:diskussion} before  Section~\ref{ch:zusammenfassung} concludes the paper and lists further open topics.

\section{Related work}\label{ch:verwandte}
The work of 
Scheller et. al. \cite{SchellerKuehnAutomated} identifies the \emph{usability of libraries} as an important quality attribute and proposes a method to evaluate it.
Surpassing this limitation to one single attribute (usability) we present a quantitative comparison based on multiple weighted attributes in the work at hand.

Wilde and Amundsen \cite{wilde_challenge_2019} describe in their work, in principle libraries can be compared and contrasted in any case.  
Their work focuses on decentralized systems rather than on libraries as we do. However, they mention that ultimately there can be no single perfect choice. We agree on that, but we want to provide a way to make a choice tailored to a given context.

Gao et. al. \cite{gao_manifold-learning_2015} addresses library recommendations in their work.
Based on a selected library, they search for other suitable libraries. 
With the help of the libraries of previous similar software projects, the authors propose alternative libraries to consider.  
Different from the work at hand, they do not use an index for their library selection. 

The work of Xie et. al. \cite{xie_multi-relation_2016} deals with an algorithm for ranking different libraries for combined use.  
The approach is not appropriate for individual library evaluation and is solely based on the library description. 
Our work proposes a method to derive a quantitative index value for individual libraries through which they can be directly compared.

The German Federal Office for Information Security (BSI) carried out a comparison of several existing cryptographic libraries \cite{BSI.2015}. 
The goal of the comparison was to find a library for further development.
The rating criteria were generated from the technical guideline BSI-TR-02102 \cite{BSI-TR-2017}. 
This comparison is very specific to the project objective. The safety aspects from BSI-TR-02102 can serve as a basis for the evaluation of security.
Our $crypto_{lib}$ index, in contrast, is customizable to any project.

In the following section, attributes for our library index are obtained from additional related work not considered in this section. 

\section{Attributes collection}
\label{ch:attribute:intro}
In order to collect attributes known to the scientific community and possibly identify new relevant ones for our library evaluation, the following two methods were used: Literature review and interviews with decision-makers.
Section~\ref{ch:attribute:literatur} describes from which scientific sources attributes were identified by literature research.
Subsequently, Section~\ref{ch:attribute:interview} describes how additional attributes were identified through interviews with decision-makers. 
These attributes confirm or complement the attributes identified from the literature review.

\subsection{Attributes identified by a literature survey}\label{ch:attribute:literatur}
The literature search was conducted in the digital publication libraries of IEEE, ACM, and Springer.
From these sources, seven papers were identified that defined or described attributes for library evaluation.
These seven papers are briefly presented in the remainder of this section.
\ifthenelse{\boolean{long}}{
The attributes described by the respective authors can be taken from \autoref{tab:allAttributes} with reference to their source.
 For example, the attribute \emph{Role Expressiveness} (ID: $1.2.1$)  is acquired from paper \cite{clarke_measuring_nodate}, which is presented next.
 }{
 }

Clarke's \emph{Cognitive Dimensions Framework} \cite{clarke_measuring_nodate} presents twelve factors by which developers are influenced when working with a library. 
The attributes were created by the authors with the use of Visual Basic classes in their minds.
All twelve attributes have been adopted by us.

The authors Zghidi et. al. extend Clarke's framework in their paper \cite{7965488} by another eleven attributes that were collected in a study in a software company that is not described further. 
In their study stakeholders brought a stronger technical perspective to the process. All eleven attributes were included in further consideration by us. 

In the work of Grill et al. \cite{grill2012methods} heuristics are used to identify usability obstacles.
The heuristics were determined by a study. All 14 attributes from this work were adopted by us.

Myers et al. \cite{myers2016improving} describe the usability of libraries using a user-centered design process. 
The work provides a platform as well as a cross-purpose guide to improving library usability and lists approaches to validating a library for this purpose. 
All ten attributes from this work were adopted by us.

Ease of learning libraries is of highest relevance for the authors of \cite{ko2004six}.
They derive so-called ``learning barriers'' from studies carried out in their work.
We derived five attributes relevant for library evaluation from these ``learning barriers'' for further use in our work.

In \cite{32713}, Bloch shows why a good library design is important for usability and how it can be considered in development. 
From this work, 21 attributes are extracted and used by us.

Using the method of categorization by \cite{lazar_research_2017}, additional attributes were identified with the work of Acar et. al. \cite{acar2017comparing}.
Although \cite{acar2017comparing} focuses on the applicability of Python libraries, they define several points that can be used for general scoring.
Five attributes were adopted by us.

The papers presented above \cite{clarke_measuring_nodate,7965488,grill2012methods,myers2016improving,ko2004six,32713,acar2017comparing} focus on the usability of libraries. 
The works predominantly pay attention to good applicability and the necessary comprehensibility of the library. 
It is noticeable that they demand extensive and detailed documentation, preferably with many examples. 
The quality of library documentation is mentioned most frequently as an attribute and thus enjoys high relevance among the attributes. 

In conclusion, the literature search produced a set of $78$ unique attributes that are considered in our further process.
Some attributes were discovered in several different works
 as is documented in 
\ifthenelse{\boolean{long}}{
\autoref{tab:allAttributes}.
}{
the long version\footnote{Long version of this work: \longVersionURL} of this work.
}

Except for \cite{acar2017comparing}, the works presented here did not consider cryptography libraries.
Accordingly, cryptography specific attributes were additionally identified through the interviews described in the following section.

\subsection{Attributes identified by interviews with decision-makers}
\label{ch:attribute:interview}
Additional attributes were identified via interviews with decision-makers. 
The interviews intended to examine the attributes found in the literature for their actuality and additionally to identify further relevant attributes specific to cryptography libraries.
 
The interviews were conducted with individuals who are in positions of making technical decisions in software projects in their professional environment.
Decision-makers such as technical directors, CTOs, or autonomous developers were explicitly invited to the interviews. 
It is important that these were people who were involved in cryptography library selection in the past or will be in the near future.

A total of $5$ interviews were conducted. 
$4$ people had already selected a cryptography library at that point. 
The average work experience is $8.1$ years. 
The people were recruited from companies and the social circle of our university. 
The interviews were conducted in compliance with the applicable General Data Protection Regulation (GDPR) and in accordance with §24 of the Hessian Data Protection and Freedom of Information Act (HDSIG)\footnote{Hessisches Datenschutz- und Informationsfreiheitsgesetz}. 
 
They were performed via different online conferencing systems. 
In the process, the audio was recorded with the consent of the subjects.
Subsequently, the recordings were anonymized by verbatim transcription.
Qualitative content analysis using a categorization method described in \cite{kuckartz_qualitative_2018} was used to extract attributes from the interviews.
Since the interviews were conducted independently from the literature review, we refrained from defining a grouping with justified evaluation aspects before categorizing.  
Therefore, categorization grouping was performed using Mayring's inductive category formation method \cite{mayring_qualitative_2000}. 

\ifthenelse{\boolean{long}}{
The interviewer followed the interview guide (Appendix~\ref{inter:qst}) to ensure consistent conditions. 
}{
The interviewer followed the interview guide (in the Appendix of the long version\footnotemark[2] of this work) to ensure consistent conditions.
}
In order to reduce initial contact inhibitions and to be able to better assess the people, demographic information and previous experience in projects were shortly discussed at the beginning of each interview.
Subsequent questions addressed the library decision-making process used in the company or project in which the people were involved.
\ifthenelse{\boolean{long}}{
Through the questions formulated in Appendix~\ref{inter:qst}, }{
Through the questions formulated in the long version\footnotemark[2] of this work, 
}
the interview aimed to find out whether there is a decision-making process for choosing libraries and how it is formulated or lived.

This allowed us to generate attributes of interest for scoring from possible existing specifications imposed on people in their projects. 
Likewise, the composition of the team or the decision-making body is interesting; such as areas of expertise and interests of the people involved. Possibly, evaluation criteria result from organizational dependencies. 

To identify more attributes, the interviewer asked how this group of people find a decision and what sources are used in this process.
At the end of the interview, the people were given the opportunity to ask further questions or to express themselves freely.

From the interviews, through a qualitative content analysis of the transcription, $50$ attributes were identified.
\ifthenelse{\boolean{long}}{
These are tagged in \autoref{tab:allAttributes} with an ``I'' for interview and the number of mentions across all interviews, for example ``(I:3)''.
}{
These are presented in the long version of this work.
}
In addition to the collection of attributes for scoring, it is obvious from the interviews that there is a strong personal opinion or preference among the people about the procedure.
These preferences come into effect differently at evaluating and making decisions about libraries.
For people without a given fixed decision-making process in the company or the project, personal opinion shapes relevant decision criteria.

\subsection{Synthesis of Section \ref{ch:attribute:literatur} and Section \ref{ch:attribute:interview}} \label{ch:attribute:synthese}

The $78$ attributes obtained from the literature review and $50$ from the interviews can be considered complementary sources, as they were collected using independent methods. 
The $128$ attributes were categorized by the authors based on their descriptions or the context in which they are placed.
In this process, the same, similar or related attributes were grouped under a meaningful attribute and structured hierarchically.
\ifthenelse{\boolean{long}}{
In \autoref{tab:allAttributes} this is represented by levels $1 - 4$ and the ID.
}{
In the long version 
of this work this is represented by levels $1 - 4$ and the ID.
}
In the further course of this work, level $1$ is referred to as attribute. 
Since the respective levels $2 - 4$ are more finely granular than the superordinate level $1$, they are suitable as evaluation criteria and will be referred to as such in the following.

\ifthenelse{\boolean{long}}{
This process resulted in $15$ attributes which are listed in \autoref{table:score:atts}.
\small
\begin{table}
	\centering
	\caption{The index attributes}
    \label{table:score:atts}
	\begin{tabular}{| p{1.5cm} | p{6cm} |}	
		\hline
		Name & Description  \\
		\hline
		Ease of Use & How much information can you extract intuitively while using?\\
		\hline
		Scalability & Is the work sequence running synchronous and parallel and what data sizes are handleable?\\
		\hline
		Testability & Is the API easily testable and debuggable, can you look up the system status at any time, are errors caught, shown, and logged?\\
		\hline
		Extendability & Is the functionality extendable and how much effort do these changes of the API make?\\
		\hline
		Functional Completeness & Is the API functionally complete, does it pack all features needed, and is it purposefully?\\
		\hline
		Data Types & What data types are being used, are they intuitive, do parameters and return values fit the functions and are they ordered consistently?\\
		\hline
		Code Quality & Does the code stick to standards and conventions?\\
		\hline
		Cost & What costs are caused by the API and which licenses are available?\\
		\hline
		Requirements & What are the requirements of the API and does the API cause dependencies that need to be solved?\\
		\hline
		Complexity & How complex is the API and how flexible can it be used and configured? What is the design aesthetic and how much boilerplate code (same code that needs to be repeated many times with no changes) needs to be written?\\
		\hline
		Maintained & How maintained is the API, meaning is it being developed further and is support being provided?\\
		\hline
		Spread & How widespread is the API and how big is the community? What are the opinions on the API, how is its reputation, are there successful stories or recommendations?\\
		\hline
		Performance Impact & How does the API impact performance and latency?\\
		\hline
		Security & Are the API and the procedures used secure?\\
		\hline
		Doc\-um\-en\-ta\-tion & Is the API documented thoroughly and are best practices being followed or examples are given?\\[1ex]
		\hline
    \end{tabular}  
\end{table}
\normalsize 
}{
This process resulted in $15$ attributes which are listed here:
\begin{enumerate}\label{table:score:atts}
	\item \label{attr:eou} \textbf{Ease of Use}: How much information can you extract intuitively while using?
		  \item \label{attr:scale} \textbf{Scalability}: Is the work sequence running synchronous and parallel and what data sizes are handleable?
		\item \label{attr:test} \textbf{Testability}: Is the API easily testable and debuggable, can you look up the system status at any time, are errors caught, shown, and logged?
		\item \label{attr:ext} \textbf{Extendability}: Is the functionality extendable and how much effort do these changes of the API make?
		\item \label{attr:fnct} \textbf{Functional Completeness}: Is the API functionally complete, does it pack all features needed, and is it purposefully?
		 \item \label{attr:types} \textbf{Data Types}: What data types are being used, are they intuitive, do parameters and return values fit the functions, and are they ordered consistently?
		\item \label{attr:quali} \textbf{Code Quality}: Does the code stick to standards and conventions?
		\item \label{attr:cost} \textbf{Cost}: What costs are caused by the API and which licenses are available?
		\item \label{attr:req} \textbf{Requirements}: What are the requirements of the API and does the API cause dependencies that need to be solved?
		\item \label{attr:compl} \textbf{Complexity}: How complex is the API and how flexible can it be used and configured? What is the design aesthetic and how much boilerplate code (same code that needs to be repeated many times with no changes) needs to be written?
		\item \label{attr:main} \textbf{Maintained}: How maintained is the API, meaning is it being developed further, and is support being provided?
		\item \label{attr:spr} \textbf{Spread}: How widespread is the API and how big is the community? What are the opinions on the API, how is its reputation, are there successful stories or recommendations?
		\item \label{attr:perf} \textbf{Performance Impact}: How does the API impact performance and latency?
		\item \label{attr:sec} \textbf{Security}: Are the API and the procedures used secure?
		\item \label{attr:doc} \textbf{Doc\-um\-en\-ta\-tion}: Is the API documented thoroughly and are best practices being followed or examples given?
		
\end{enumerate}
}
 
The description of the respective attributes listed there includes an explanation of the term.
These were generated from the description of the attributes in the literature or the contexts in the interviews.
The description of the attributes is intended to help developers later when weighting the attributes.

\subsection{Validation of the synthesis via survey} \label{ch:attribute:validierung}
By an online survey, the results
\ifthenelse{\boolean{long}}{
in \autoref{table:score:atts} 
}
were confirmed.
Participation in the survey was voluntary and in compliance with the currently applicable data protection regulations GDPR and §24 HDSIG. 
People acquisition was conducted via direct messaging, social media, and email.
A total of $36$ people participated in the survey.
Twelve people performed them completely.
\ifthenelse{\boolean{long}}{
Appendix \ref{appendix:fragebogen} lists the questionnaire.
}{
The long version
of this work lists the questionnaire used.
}
The average programming experience of the people is $7.88$ years.
\ifthenelse{\boolean{long}}{
Figure~\ref{abb:survey:oft} shows the frequency with which people make decisions about libraries. 
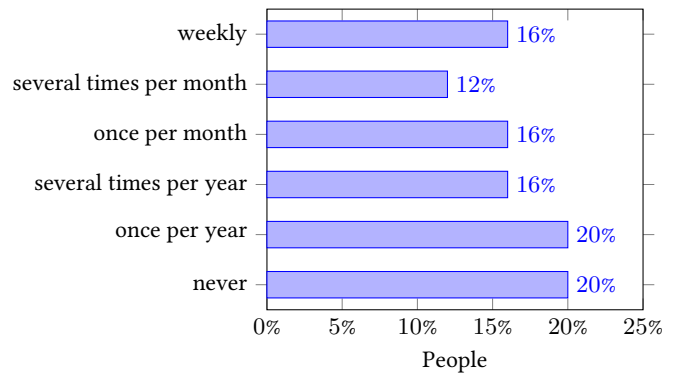
\begin{figure}
	\centering
	\caption{Evaluation of the question \ref{qst:qst:2}: How often do you select an API?}
	\label{abb:survey:oft}
	\begin{tikzpicture}
		\begin{axis}[
			scale only axis=true,
			width=5cm,
			height=4cm,
			xbar,
			legend style={at={(0.5,-0.15)},
				anchor=north,legend columns=-1},
			xlabel={People},
			xmin=0,
			xmax=0.25, 
			xticklabel={\pgfmathparse{\tick*100}\pgfmathprintnumber{\pgfmathresult}\%},
			symbolic y coords={never,once per year,several times per year,once per month,several times per month,weekly},
			ytick=data,
			point meta={x*100},
			nodes near coords={\pgfmathprintnumber\pgfplotspointmeta\%},
			nodes near coords align={horizontal},
			]
			\addplot coordinates {(0.20,never)(0.20,once per year)(0.16,several times per year)(0.16,once per month)(0.12,several times per month)(0.16,weekly)};
		\end{axis}
	\end{tikzpicture}
\end{figure}
$24\%$ of the people have already selected a cryptography library.
}

\pgfplotstableread[col sep=semicolon,header=true]{
	A; aFehler; aLabel; B; bFehler; bLabel; Frage
4.277777778;0.772579631; {(mean: 4,28; SD: 0,77)};4.384615385;0.938723409; {(mean: 4,38; SD: 0,94)};Ease of Use
3.944444444;0.694012211; {(mean: 3,94; SD: 0,69)};4.230769231;0.896054163; {(mean: 4,23; SD: 0,9)};Scalability
4.5;0.824957911; {(mean: 4,5; SD: 0,82)};4.461538462;0.960058032; {(mean: 4,46; SD: 0,96)};Testability
4.166666667;0.746390491; {(mean: 4,17; SD: 0,75)};4.230769231;0.896054163; {(mean: 4,23; SD: 0,9)};Extendability
4.0625;0.765625; {(mean: 4,06; SD: 0,77)};4.384615385;0.938723409; {(mean: 4,38; SD: 0,94)};Functional Completeness
4.1875;0.796875; {(mean: 4,19; SD: 0,8)};4.307692308;0.917388786; {(mean: 4,31; SD: 0,92)};Data Types
4.5;0.875; {(mean: 4,5; SD: 0,88)};4.083333333;0.890081665; {(mean: 4,08; SD: 0,89)};Code Quality
4.25;0.8125; {(mean: 4,25; SD: 0,81)};4.076923077;0.853384917; {(mean: 4,08; SD: 0,85)};Cost
4.117647059;0.756140478; {(mean: 4,12; SD: 0,76)};4.307692308;0.917388786; {(mean: 4,31; SD: 0,92)};Requirements
3.733333333;0.705743632; {(mean: 3,73; SD: 0,71)};3.923076923;0.810715671; {(mean: 3,92; SD: 0,81)};Complexity
4.529411765;0.856008088; {(mean: 4,53; SD: 0,86)};4.230769231;0.896054163; {(mean: 4,23; SD: 0,9)};Maintained
4.294117647;0.798940882; {(mean: 4,29; SD: 0,8)};3.846153846;0.789381048; {(mean: 3,85; SD: 0,79)};Spread
4.117647059;0.756140478; {(mean: 4,12; SD: 0,76)};4.538461538;0.981392655; {(mean: 4,54; SD: 0,98)};Performance Impact
4.5;0.875; {(mean: 4,5; SD: 0,88)};4.615384615;1.002727278; {(mean: 4,62; SD: 1,00)};Security
4.647058824;0.884541691; {(mean: 4,65; SD: 0,88)};4.384615385;0.938723409; {(mean: 4,38; SD: 0,94)};Documentation
}\data

\ifthenelse{\boolean{long}}{
\begin{figure}
	\centering
		\caption{Evaluation of question \ref{qst:qst:4} in blue: "Does the description match the term?" and question~\ref{qst:qst:5} in red: "Is the term suitable for the rating of an API (not in context of own projects)?" with mean and standard deviation (SD).}
	\label{abb:survey:rohr}
\begin{tikzpicture}
	
	\begin{axis}[ nodes near coords,
		grid=major,
		height= 15cm,
		width= 7cm,
		xtick={1,2,...,5},
		xtick pos=top,
		xmin = 2.8,
		xmax = 5.8,
		ytick = data,
		yticklabels from table = {\data}{Frage},
		y dir=reverse,
		ytick align=center,
		ytick pos=left,
		yticklabel style={text width=2.5cm}, 
		axis line style={-},
		legend style={at={(0.15,1.15)} },
		xticklabels={$(1)$ not at all,$(2)$ hardly,$(3)$ moderately,$(4)$ fairly,$(5)$ extraordinarily},
		xticklabel style={rotate=45,anchor=west,yshift=1mm,xshift=-1mm}
		]
		
		\addplot+[only marks, every mark/.append style={yshift=1mm},
		point meta=explicit symbolic, 
		nodes near coords style={font=\tiny,anchor=south,yshift=-0.5mm},
		error bars/.cd, error bar style={yshift=1mm},
		x dir=both,
		x explicit,
		] 
		table[y expr =\coordindex, x expr={\thisrow{A}},x error expr={\thisrow{aFehler}},meta=aLabel] \data;
		
		\addplot+[only marks, every mark/.append style={yshift=-1mm},
		point meta=explicit symbolic, 
		nodes near coords style={font=\tiny,anchor=north,yshift=0.2mm},
		error bars/.cd, error bar style={yshift=-1mm},
		x dir=both,
		x explicit,
		] 
		table[y expr =\coordindex, x expr={\thisrow{B}},x error expr={\thisrow{bFehler}},meta=bLabel] \data;

	\end{axis}
\end{tikzpicture}
\end{figure}
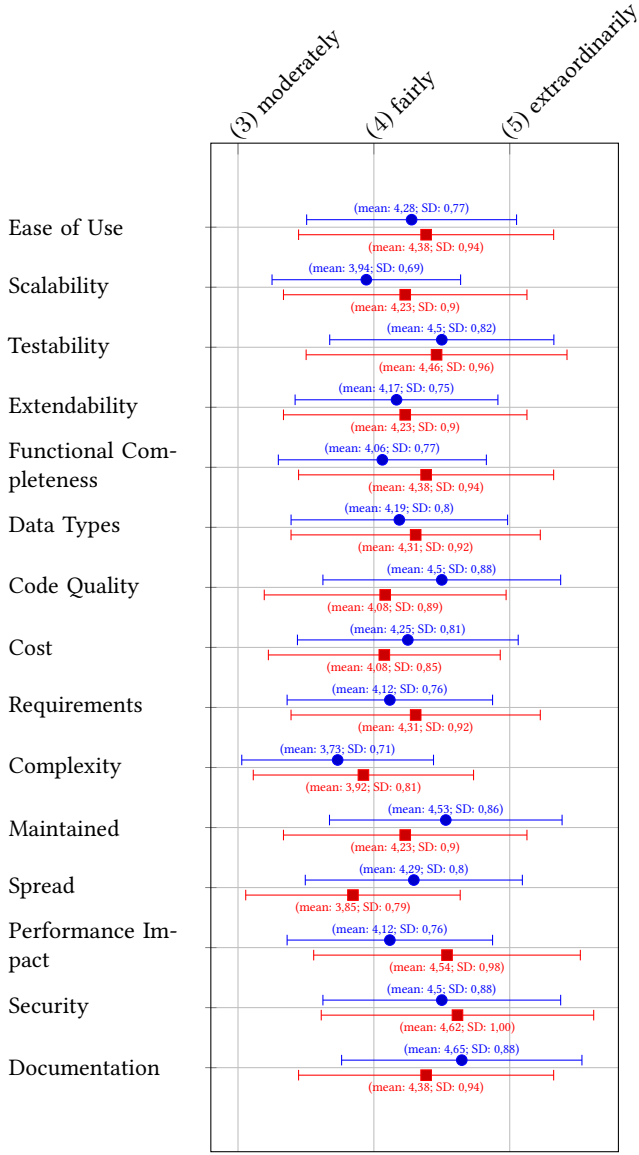
}{
\begin{figure}
	\centering
		\caption{Evaluation of question~4 in blue: "Does the description match the term?" and question~5 in red: "Is the term suitable for the rating of an API (not in context of own projects)?"; both with mean and standard deviation (SD). $(3)$ moderately,$(4)$ fairly, $(5)$ extraordinarily}
	\label{abb:survey:rohr}
\begin{tikzpicture}
	
	\begin{axis}[ nodes near coords,
		grid=major,
		height= 15cm,
		width= 7cm,
		xtick={1,2,...,5},
		xtick pos=top,
		xmin = 2.8,
		xmax = 5.8,
		ytick = data,
		yticklabels from table = {\data}{Frage},
		y dir=reverse,
		ytick align=center,
		ytick pos=left,
		yticklabel style={text width=2.5cm}, 
		axis line style={-},
		legend style={at={(0.15,1.15)} },
		xticklabels={1,2,3,4,5}
		]
		
		\addplot+[only marks, every mark/.append style={yshift=1mm},
		point meta=explicit symbolic, 
		nodes near coords style={font=\tiny,anchor=south,yshift=-0.5mm},
		error bars/.cd, error bar style={yshift=1mm},
		x dir=both,
		x explicit,
		] 
		table[y expr =\coordindex, x expr={\thisrow{A}},x error expr={\thisrow{aFehler}},meta=aLabel] \data;
		
		\addplot+[only marks, every mark/.append style={yshift=-1mm},
		point meta=explicit symbolic, 
		nodes near coords style={font=\tiny,anchor=north,yshift=0.2mm},
		error bars/.cd, error bar style={yshift=-1mm},
		x dir=both,
		x explicit,
		] 
		table[y expr =\coordindex, x expr={\thisrow{B}},x error expr={\thisrow{bFehler}},meta=bLabel] \data;

	\end{axis}
\end{tikzpicture}
\end{figure}
}

In Figure \ref{abb:survey:rohr}, the questions
\ifthenelse{\boolean{long}}{
\ref{qst:qst:4} and \ref{qst:qst:5} 
}{
4 and 5 
}
are answered by people using a five-point scale (not at all = $1$, hardly = $2$, moderately = $3$, fairly = $4$, extraordinarily = $5$ \cite{rohr1978skalen}).
It can be seen from the blue dots (question no.
\ifthenelse{\boolean{long}}{
\ref{qst:qst:4})
}{
4)
}
 that all descriptions were rated by them as matching the attributes with a tendency to \emph{fairly}.

In question 
\ifthenelse{\boolean{long}}{
no.~\ref{qst:qst:5},
}{
no.~5,
}
people were asked whether they felt the attributes were suitable for an index.
All attributes were rated as tending to be \emph{fairly} suitable for an index. 
In Figure \ref{abb:survey:rohr} this can be seen by the red squares. 

With the question 
\ifthenelse{\boolean{long}}{
\ref{qst:qst:6},
}{
6,
}
the people were asked to rate the relevance of the attributes according to their personal assessment based on a sorting of the attributes. 
Rank 15 means that this attribute is the most relevant.
The lower the rank, the less important the attribute.
In \autoref{table:survey:sort}, people rankings were evaluated by the mean rank method \cite{hedderich2016angewandte}. The attributes can be placed in multiple rows by varying people's votes. Number of votes in brackets.
\small
\begin{table}
	\centering
	\captionof{table}{Evaluation of the question 6: personal assessment of the relevance of the attributes. (Sorted in descending order from Rank $15 =$ as most important, to Rank $1 =$ least important).}
	\label{table:survey:sort}
	\begin{tabular}{| c | p{7.4cm} |}
		\hline
		Ra & Attribute (Elected by x people to this rank)\\ [0.5ex]
		\hline
		 15 & Ease of Use~(4); Security~(4); Functional Completeness~(1); Code Quality~(1); Requirements~(1); Documentation~(1)\\
		\hline
		 14 & Documentation~(3); Functional Completeness~(2); Security~(2); Ease of Use~(1); Scalability~(1); Testability~(1); Extendability~(1); Spread~(1)\\
		\hline
		 13  & Maintained~(2); Security~(2); Scalability~(1); Extendability~(1); Functional Completeness~(1); Code Quality~(1); Requirements~(1); Spread~(1); Performance Impact~(1); Documentation~(1)\\
		\hline
		 12 & Functional Completeness~(3); Documentation~(3); Ease of Use~(1); Scalability~(1); Testability~(1); Extendability~(1); Code Quality~(1); Requirements~(1)\\
		\hline
		 11 & Testability~(3); Functional Completeness~(2); Ease of Use~(1); Extendability~(1); Code Quality~(1); Requirements~(1); Maintained~(1); Performance Impact~(1); Security~(1)\\
		\hline
		 10 & Maintained~(3); Requirements~(2); Complexity~(2); Ease of Use~(1); Scalability~(1); Cost~(1); Spread~(1); Performance Impact~(1)\\
		\hline
		 9 & Extendability~(2); Code Quality~(2); Cost~(2); Scalability~(1); Testability~(1); Data Types~(1); Requirements~(1); Complexity~(1); Performance Impact~(1)\\
		\hline
		 8 & Cost~(3); Testability~(2); Performance Impact~(2); Extendability~(1); Data Types~(1); Code Quality~(1); Complexity~(1); Spread~(1)\\
		\hline
		 7 & Scalability~(2); Complexity~(2); Security~(2); Ease of Use~(1); Extendability~(1); Functional Completeness~(1); Requirements~(1); Spread~(1); Performance Impact~(1)\\
		\hline
		 6 & Maintained~(3); Scalability~(2); Code Quality~(2); Ease of Use~(1); Functional Completeness~(1); Cost~(1); Complexity~(1); Performance Impact~(1)\\
		\hline
		 5 & Requirements~(3); Ease of Use~(1); Testability~(1); Data Types~(1); Cost~(1); Complexity~(1); Maintained~(1); Spread~(1); Security~(1); Documentation~(1)\\
		\hline
		 4 & Complexity~(4); Performance Impact~(2); Documentation~(2); Scalability~(1); Testability~(1); Data Types~(1); Maintained~(1)\\
		\hline
		 3 & Data Types~(3); Extendability~(2); Cost~(2); Spread~(2); Performance Impact~(2); Code Quality~(1)\\
		\hline
		 2 & Data Types~(4); Testability~(2); Spread~(2); Ease of Use~(1); Cost~(1); Maintained~(1); Documentation~(1)\\
		\hline
		 1 & Spread~(2); Code Quality~(2); Scalability~(1); Extendability~(1); Functional Completeness~(1); Data Types~(1); Cost~(1); Requirements~(1)\\ [0.5ex]
		\hline
	\end{tabular}
\end{table}
\normalsize 

From these results, a preference of the people is deduced.  
It becomes clear that \emph{Ease of Use} and \emph{Security} are the most relevant attributes for cryptography libraries from the people's point of view. For both attributes received the most votes (four) in 15th place.
Using the preferences from \autoref{table:survey:sort}, a reference weighting for the index is derived in the next section.

\section{Index definition} \label{ch:definition}
In order to create a comparable consistent weighting for the index, the number of mentions of the attributes from the literature, the interviews and the weighting from \autoref{table:survey:sort} are put together.
The number of mentions in the literature, the interviews, and the rankings from \autoref{table:survey:sort} are each normalized by the rank method~\cite{hedderich2016angewandte}.
The average of these three sources (column no. VI in \autoref{tab:score:weights}) is then ranked again (in column no. VII). Based on the rank in column no. VII, the possible values for ($g_i$) (column no. VIII in \autoref{tab:score:weights}) $1.5; 1.25; 1; 0.75; 0.5$ are assigned as following: the ranks 15 - 13 get 1.5 points each, the ranks 12 - 10 get 1.25 points each and so on.
This procedure is shown in \autoref{tab:score:weights} with the intermediate results and yields the reference weighting ($g_i$) of the attributes in column~no.~VIII.

The weighting of the attributes can be chosen differently for each software project.
This is realized by individual adjustments to the weighting in \autoref{tab:score:weights} column no. VIII. 
The sum of the individual weights should correspond to the number of attributes.
This is to prevent excessive weighting.

\newcolumntype{L}[1]{>{\raggedright\arraybackslash}p{#1}} 
\newcolumntype{C}[1]{>{\centering\arraybackslash}p{#1}} 
\newcolumntype{R}[1]{>{\raggedleft\arraybackslash}p{#1}} 
\ifthenelse{\boolean{long}}{
\begin{table*}
	\caption{The calculation of the reference weight ($g_i$) for the index. The number of mentions is listed in the brackets as the basis for the rank.}
	\label{tab:score:weights}
	\centering
	\begin{tabular}{|C{0.3cm}| L{2.6cm} | C{2cm} | C{2cm} | C{2cm} | C{1cm} | C{1.35cm} | C{1.35cm} |}
	    \hline
	    I & \multicolumn{1}{c|}{II} & III & IV & V & VI & VII & VIII \\
		\hline
		No & Attribut & Rank Literature (number of mentions) & Rank Interviews (number of mentions) & Rank Questionnaire (\autoref{table:survey:sort}) & $\oslash$ & Rank Total & Weight ($g_i$)\\
		\hline
		\ref{attr:eou}& Ease of Use & 15 (33) & 12 (10) & 15 & 14 & 15 & 1,5\\
		\hline
		\ref{attr:scale}& Scalability & 6,5 (2) & 8 (4) & 9 & 7,83 & 5,5 & 0,75\\
		\hline
		\ref{attr:test}& Testability & 14 (9) & 5 (1) & 11 & 10 & 12 & 1,25\\
		\hline
		\ref{attr:ext}& Extendability & 8 (3) & 2,5 (0) & 10 & 6,83 & 3 & 0,5\\
		\hline
		\ref{attr:fnct}& Functional Completeness & 10,5 (4) & 2,5 (0) & 12 & 8,33 & 7 & 1,0\\
		\hline
		\ref{attr:types}& Data Types & 10,5 (4) & 2,5 (0) & 7 & 7,67 & 4 & 0,75\\
		\hline
		\ref{attr:quali}& Code Quality & 10,5 (4) & 10 (5) & 10 & 10,17 & 13 & 1,5\\
		\hline
		\ref{attr:cost}& Cost & 2,5 (0) & 13 (11) & 10 & 8,5 & 8 & 1,0\\
		\hline
		\ref{attr:req}& Requirements & 2,5 (0) & 8 (4) & 8 & 6,17 & 1 & 0,5\\
		\hline
		\ref{attr:compl}& Complexity & 13 (8) & 6 (3) & 9 & 9,33 & 10,5 & 1,25\\
		\hline
		\ref{attr:main}& Maintained & 2,5 (0) & 11 (8) & 13 & 8,83 & 9 & 1,0\\
		\hline
		\ref{attr:spr}& Spread & 2,5 (0) & 15 (26) & 6 & 7,83 & 5,5 & 0,75\\
		\hline
		\ref{attr:perf}& Performance Impact & 6,5 (2) & 2,5 (6) & 11 & 6,67 & 2 & 0,5\\
		\hline
		\ref{attr:sec}& Security & 5 (1)& 8 (4)& 15 & 9,33 & 10,5 & 1,25\\
		\hline
		\ref{attr:doc}& Do\-cu\-men\-ta\-tion & 10,5 (4)& 14 (19)& 14 & 12,83 & 14 & 1,5\\
		\hline
		\multicolumn{7}{c|}{} & $ \sum $ 15 \\
		\cline{8-8}
	\end{tabular}
\end{table*}
}{ 
\small
\begin{table}
	\caption{The calculation of the reference weight ($g_i$) for the index. The number of mentions is listed in the brackets as the basis for the rank.}
	\label{tab:score:weights}
	\centering
	\begin{tabular}{|C{0.12cm}| L{1.4cm} | C{1cm} | C{1cm} | C{0.8cm} | C{0.4cm} | C{0.5cm} | C{0.55cm} |}
	    \hline
	    I & \multicolumn{1}{c|}{II} & III & IV & V & VI & VII & VIII \\
		\hline
		\# & Attribute & Rank Literature (number of mentions) & Rank Interviews (number of mentions) & Rank Questionnaire Tab. \ref{table:survey:sort} & $\oslash$ & Rank Total & Wei\-ght ($g_i$)\\
		\hline
		\ref{attr:eou}& Ease of Use & 15 (33) & 12 (10) & 15 & 14 & 15 & 1,5\\
		\hline
		\ref{attr:scale}& Scalability & 6,5 (2) & 8 (4) & 9 & 7,83 & 5,5 & 0,75\\
		\hline
		\ref{attr:test}& Testability & 14 (9) & 5 (1) & 11 & 10 & 12 & 1,25\\
		\hline
		\ref{attr:ext}& Extendability & 8 (3) & 2,5 (0) & 10 & 6,83 & 3 & 0,5\\
		\hline
		\ref{attr:fnct}& Functional Complet. & 10,5 (4) & 2,5 (0) & 12 & 8,33 & 7 & 1,0\\
		\hline
		\ref{attr:types}& Data Types & 10,5 (4) & 2,5 (0) & 7 & 7,67 & 4 & 0,75\\
		\hline
		\ref{attr:quali}& Code Quality & 10,5 (4) & 10 (5) & 10 & 10,17 & 13 & 1,5\\
		\hline
		\ref{attr:cost}& Cost & 2,5 (0) & 13 (11) & 10 & 8,5 & 8 & 1,0\\
		\hline
		\ref{attr:req}& Requirements & 2,5 (0) & 8 (4) & 8 & 6,17 & 1 & 0,5\\
		\hline
		\ref{attr:compl}& Complexity & 13 (8) & 6 (3) & 9 & 9,33 & 10,5 & 1,25\\
		\hline
		\ref{attr:main}& Maintained & 2,5 (0) & 11 (8) & 13 & 8,83 & 9 & 1,0\\
		\hline
		\ref{attr:spr}& Spread & 2,5 (0) & 15 (26) & 6 & 7,83 & 5,5 & 0,75\\
		\hline
		\ref{attr:perf}& Performance Impact & 6,5 (2) & 2,5 (6) & 11 & 6,67 & 2 & 0,5\\
		\hline
		\ref{attr:sec}& Security & 5 (1)& 8 (4)& 15 & 9,33 & 10,5 & 1,25\\
		\hline
		\ref{attr:doc}& Do\-cu\-men\-ta\-tion & 10,5 (4)& 14 (19)& 14 & 12,83 & 14 & 1,5\\
		\hline
		\multicolumn{7}{c|}{} & $ \sum $ 15 \\
		\cline{8-8}
	\end{tabular}
\end{table}
\normalsize 
}
The $crypto_{lib}$ index of a cryptography library is calculated from the sum of all evaluation criteria of an attribute divided by the number of evaluation criteria.
This sum is multiplied by the weight of the attribute and the sum of all attributes gives the $crypto_{lib}$ index.
Thus, the $crypto_{lib}$ index is defined as follows: 
$$
crypto_{lib} \mbox{ index} = \sum_{i = 1}^{n} \frac{\sum_{j = 1}^{m_i} b_{ij}}{m_i}  \cdot g_i
$$
Where $b_{ij}$ stands for evaluation criterion $j$ of attribute $i$;
$m_i$~=~number of $b_j$ in attribute $i$;
$ n $ = number of attributes
 and $g_i$ stands for the weighting of the respective attribute $i$ (see \autoref{tab:score:weights} column~no.~VIII).

We construct the formula in this way because, this index treats attributes with many evaluation criteria, the same as attributes with few once. 
This allows evaluation criteria to be easily added (or omitted) in a possible new version of this index. 
Multiple evaluation criteria for a single attribute ensure that the evaluation of the attribute will be more fine-grained and accurate.
The weighting of the respective attribute has more influence on the $crypto_{lib}$ index than the number of evaluation criteria of an attribute.
This makes the index flexibly adaptable to the needs of software projects.

\section{Definition of the $crypto_{lib}$ index evaluation criteria}\label{ch:index:beschreibung}
The evaluation criteria defined in
\ifthenelse{\boolean{long}}{
Appendix~\ref{app:bewertungskriterien}
}{
the long version\footnote{Long version of this work: \longVersionURL} of this work 
}
are used for library assessments.
The assessment results are used to create the $crypto_{lib}$ index described in the previous section.
In general, all evaluation criteria are normalized to the scale $[-2, -1, 0, +1, +2]$. 
For many evaluation criteria, the scale is not fully indicated within its textual description. 
In such cases, the intermediate values can be interpolated linearly.
Evaluation criteria without explicit description of normalization are presented as follows via numerical values in percent.  
This rating is referred to as the \textit{default rating} in the further text.
\begin{description}
	\item[+2] if at least 90\% of the condition of the evaluation criterion is met.
	\item[+1] if at least 75\% of the condition is met.
	\item[ 0] if at least 50\% of the condition is met.
	\item[-1] if at least 25\% of the condition is met.
	\item[-2] if less than 25\% of the condition is met.
\end{description}
The evaluation criteria are listed and described in detail in \ifthenelse{\boolean{long}}{
Appendix~\ref{app:bewertungskriterien}. 
}{
the long version\footnotemark[4] of this work.
}

\section{Exemplary use} \label{ch:index:beispiel}
In the previous section, the evaluation criteria of the $15$ attributes were introduced. 
These were used to evaluate the Java versions of Google's Tink\footnote{Version 1.6.1, \url{https://github.com/google/tink} Retrieved 28 Oct 2021} library and the Bouncy Castle\footnote{Tag r1rv69, \url{https://github.com/bcgit/bc-java} Retrieved 28 Oct 2021} library. 
The two libraries were chosen because they appeal to the similar target group of developers.
They are both developed and maintained by an open source community. 
Bouncy Castle is established. 
Tink is relatively new (since 2017) and claims to be easy to use for developers without cryptographic knowledge\footnote{\url{https://developers.google.com/tink} Retrieved 22 Oct 2021}. 

\ifthenelse{\boolean{long}}{
\begin{table}
	\caption{Evaluation criteria (withe background), Attributes (gray background)  and resulting $crypto_{lib}$ index of the libraries Bouncy Castle and Tink.}
	\centering
  \label{tab:bewertunglong}
  \begin{tabular}{|l l l l|}
  \hline
  Nr & Attribute     & Bouncy Castle      & Tink  \\ 
  \hline
  \rowcolor[gray]{0.9} \ref{attr:eou}  & \textbf{Ease of Use}  & $+0.33$ & $+1$ \\
  \ref{attr:eou:read} & Readability & $0$ & $-1$ \\
  \ref{attr:eou:defaults} & Default Settings & $0$ & $+2$ \\
  \ref{attr:eou:naming} & Naming Conventions & $+1$ & $+2$ \\
  \hline
  \rowcolor[gray]{0.9}\ref{attr:scale}  & \textbf{Scalability}  & $0$ & $0$ \\
  \ref{attr:scale:para} & Concurrency & $0$ & $0$ \\
  \hline
  \rowcolor[gray]{0.9}\ref{attr:test} & \textbf{Testability}  & $+1$ & $+0.5$ \\
  \ref{attr:test:test} & Testability & $0$ & $0$\\
  \ref{attr:test:exc} & Exceptions & $+2$ & $+1$ \\
  \hline
  \rowcolor[gray]{0.9}\ref{attr:ext} & \textbf{Extendability} & $+2$ & $+2$ \\
  \ref{attr:ext:pub} & Public & $+2$ & $+2$ \\
  \ref{attr:ext:int} & Interfaces & $+2$ & $+2$ \\
  \hline
  \rowcolor[gray]{0.9}\ref{attr:fnct}  & \textbf{Functional Completeness} & $+1$ & $+2$ \\
  \ref{attr:fnct:goal} & Purposefulness & $+1$ & $+2$ \\
  \hline
  \rowcolor[gray]{0.9}\ref{attr:types}  & \textbf{Data Types}   & $+2$ & $+2$ \\
  \ref{attr:test:ret} & Returnvalues & $+2$ & $+2$ \\
  \ref{attr:types:order} & Ordering & $+2$ & $+2$ \\
  \hline
  \rowcolor[gray]{0.9}\ref{attr:quali}  & \textbf{Code Quality} & $-0.67$ & $+1.67$ \\
  \ref{attr:quali:bugs} & Bugs & $+2$ & $+1$ \\
  \ref{attr:quali:vulner} & Vulnerability & $-2$ & $+2$ \\
  \ref{attr:quali:smell} & Code Smell & $-2$ & $+2$ \\
  \hline
  \rowcolor[gray]{0.9}\ref{attr:cost}  & \textbf{Cost}         & $+2$ & $+2$ \\
  \ref{attr:cost:cost} & Cost & $+2$ & $+2$ \\
  \ref{attr:cost:lic} & Licence & $+2$ & $+2$ \\
  \hline
  \rowcolor[gray]{0.9}\ref{attr:req}  & \textbf{Requirements} & $+1$ & $+2$ \\
  \ref{attr:req:pre} & Dependencies & $+1$ & $+2$ \\
  \hline
  \rowcolor[gray]{0.9}\ref{attr:compl}  & \textbf{Complexity}  & $+0.5$ & $+1$ \\
  \ref{attr:compl:gran} & Atomic Setting & $+1$ & $0$ \\
  \ref{attr:compl:boil} & Boilerplatecode & $0$ & $+2$ \\
  \hline
  \rowcolor[gray]{0.9}\ref{attr:main}  & \textbf{Maintained}  & $0.33$ & $0$ \\
  \ref{attr:main:freq} & Release-frequency & $+1$ & $+1$ \\
  \ref{attr:main:patch} & Patch frequency & $-1$ & $0$ \\
  \ref{attr:main:sup} & Support & $+1$ & $-1$ \\
  \hline
  \rowcolor[gray]{0.9}\ref{attr:spr} & \textbf{Spread}       & $+1$ & $+0.5$ \\
  \ref{attr:spr:succ} & Successful Stories & $+1$ & $-1$ \\
  \ref{attr:spr:git} & Repositories & $+1$ & $+2$ \\ 
  \hline
  \rowcolor[gray]{0.9}\ref{attr:perf} & \textbf{Performance Impact} & $-$ & $-$ \\
  \hline
  \rowcolor[gray]{0.9}\ref{attr:sec} & \textbf{Security}     & $-0.5$ & $0$\\
  \ref{attr:sec:std} & Standards & $-1$ & $+2$ \\
  \ref{attr:sec:cert} & Certificated & $0$ & $-2$ \\
  \hline
  \rowcolor[gray]{0.9}\ref{attr:doc} & \textbf{Documentation} & $-0.5$ & $+2$ \\
  \ref{attr:doc:fkt} & Function documentation & $+1$ & $+2$ \\
  \ref{attr:doc:ex} & Examples & $-2$ & $+2$ \\
  
  \hline
  \rowcolor[gray]{0.9} & $crypto_{lib}$ index & $7.08$ & $16.75$ \\
  \hline
  \end{tabular}
\end{table}
}{
\small
\begin{table}
	\caption[]{Attributes and resulting $crypto_{lib}$ index of the libraries Bouncy Castle and Tink. The scored evaluation criteria can be found in the long version of this work.}
  \label{tab:bewertunglong}	
  \centering
  \begin{tabular}{|l l l l|}
  \hline
  Nr & Attribute     & Bouncy Castle      & Tink  \\ 
  \hline
  \ref{attr:eou}  & Ease of Use  & $+0.33$ & $+1$ \\
  \ref{attr:scale}  & Scalability  & $0$ & $0$ \\
  \ref{attr:test} & Testability  & $+1$ & $+0.5$ \\
  \ref{attr:ext} & Extendability & $+2$ & $+2$ \\
  \ref{attr:fnct}  & Functional Completeness & $+1$ & $+2$ \\
 \ref{attr:types}  & Data Types   & $+2$ & $+2$ \\
 \ref{attr:quali}  & Code Quality & $-0.67$ & $+1.67$ \\
\ref{attr:cost}  & Cost         & $+2$ & $+2$ \\
  \ref{attr:req}  & Requirements & $+1$ & $+2$ \\
  \ref{attr:compl}  & Complexity  & $+0.5$ & $+1$ \\
  \ref{attr:main}  & Maintained  & $0.33$ & $0$ \\
  \ref{attr:spr} & Spread       & $+1$ & $+0.5$ \\
  \ref{attr:perf} & Performance Impact & $-$ & $-$ \\
  \ref{attr:sec} & Security     & $-0.5$ & $0$\\
 \ref{attr:doc} & Documentation & $-0.5$ & $+2$ \\
  \hline
   & $crypto_{lib}$ index & $7.08$ & $16.75$ \\
  \hline
  \end{tabular}
\end{table}
\normalsize 
}
\ifthenelse{\boolean{long}}{
\autoref{tab:bewertunglong} shows the individual points achieved for the evaluation criteria of both libraries.
}{
\autoref{tab:bewertunglong} shows the individual points achieved for the attributes of both libraries.} 
The attributes are the average of the evaluation criteria.
\ifthenelse{\boolean{long}}{
}{
The scored evaluation criteria can be found in the long version
of this work.
}
These values were multiplied with the reference weighting from \autoref{tab:score:weights} column no. VIII and the results summed up resulting in the index value for each. 
The $crypto_{lib}$ index of Bouncy Castle is $7.08$.
Tink reaches an index of $16.75$.
With the reference weighting used, a minimum index of $-29$ points and a maximum index of $29$ points is achievable. 
\ifthenelse{\boolean{long}}{
Two detailed assessments are shown below as examples. They are based on the respective criteria in Annex \ref{app:bewertungskriterien}. 
For the evaluation criterion \ref{attr:eou:defaults} \emph{Default Settings}, the Bouncy Castle library was given a score of $0$, since obsolete algorithms can be used due to the flexible constructors.
One example is the constructor for the base AES keygen KeyGen (int) where the user can input any value. 
The Tink library has been assigned a score of $+2$, because safe default values are used. 
For the evaluation criterion \ref{attr:main:freq}  \emph{Release-frequency}, the Bouncy Castle library was given a score of $+1$, because the current\footnote{27 Jan 2021} release cycle is every four months on average.
The Tink library has achieved a rating of $+1$ as well because currently\footnotemark[6] releases are being released quarterly.

These examples, and}{
}
\autoref{tab:bewertunglong} shows that the selected evaluation criteria are applicable. 
With the reference weighting generated from the online survey preferences in Section~\ref{ch:attribute:validierung}, the Tink library gets a better $crypto_{lib}$ index. 
When compared in detail, the Tink library performs better in each of the three attributes weighted highest at times $1.5$: \emph{Ease of Use}, \emph{Code Quality}, and \emph{Documentation}. 
Moreover, with the exception for \emph{Testability}, this is the same case for the attributes with $1.25$-fold weights \emph{Complexity} and \emph{Security}.
Thus, according to the reference weighting, the Tink library is preferable to the Bouncy Castle library.

Since each software project has different conditions and requirements, the weighting can be adjusted individually. 
This leads to $crypto_{lib}$ index results adapted to the corresponding software project.

\section{Discussion and limitations} \label{ch:diskussion}
It is generally difficult to get experts for interviews. Since only 5 people participated in the interviews, it is possible that not all relevant properties for cryptographic libraries were mentioned.
Due to technical advancements, some of the evaluation criteria might change their relevance.
They should periodically be reviewed, therefore, to ensure that they are up to date. New criteria should be added to a new version of the $crypto_{lib}$ index, if necessary. 
For example, the authors are not aware of any useful evaluation criterion applicable for the attribute \emph{Performance Impact}. This might change over time.
In the same way, individual evaluation criteria show potential for improvement. 
For example, the evaluation criterion
\ifthenelse{\boolean{long}}{
\ref{attr:spr:git}~\emph{Repositories}
}{
12b~\emph{Repositories}
}
 is difficult to implement for commercial projects, since these are rarely developed on a public repository.
Other evaluation criteria are better suited for libraries written in object-oriented languages.
For example, the criteria
\ifthenelse{\boolean{long}}{
\ref{attr:test:exc}~\emph{Exceptions}, \ref{attr:ext:pub}~\emph{Public}, and \ref{attr:ext:int}~\emph{Interfaces}.
}{
3b~\emph{Exceptions}, 4a~\emph{Public}, and 4b~\emph{Interfaces}.
}
This represents a bias against procedural languages. 

Overall, the $crypto_{lib}$ index described in this work is a first step to compare cryptography libraries with each other. 
The individually adaptable weighting supports the project-specific customization to meet the needs of the software project.
The reference weighting provided in the work at hand enables libraries to be compared independently of specific projects, leading to a general quantitative comparison of crypto libraries.

\section{Conclusion and outlook}\label{ch:zusammenfassung}
This is a first attempt to create a $crypto_{lib}$ index based on literature and interviews.
By evaluating the two libraries Bouncy Castle ($crypto_{lib}$ index of $7.08$) and Tink ($crypto_{lib}$ index of $16.75$) as examples, we have shown that this index is applicable.

Our reference weighting emphasizes that \emph{Ease of Use}, \emph{Code Quality}, and \emph{Documentation} are the most relevant attributes. 
We have seen that some evaluation criteria do not equally fit to all programming languages. Furthermore, it is not always easy or even possible to find the information needed to assess the evaluation criteria.
More surveys may provide new attributes or new evaluation criteria of existing attributes for our index. 

Although it takes effort to evaluate many libraries using our index, in the end there will be a benefit for software quality. 
We envision an open access database where the $crypto_{lib}$ index of many libraries is stored and actively maintained by a committee of experts.
The ranking of individual libraries based on the reference weighting could be visualized in a graph over time to show the evolution of each library.
Decision-makers could then just adapt the weightings of the attributes to fit their particular project.



\begin{acks}
 This research work has been funded by the German Federal Ministry of Education and Research and the Hessian Ministry of Higher Education, Research, Science and the Arts within their joint support of the National Research Center for Applied Cybersecurity ATHENE.
\end{acks}

\bibliographystyle{ACM-Reference-Format}

\ifthenelse{\boolean{long}}{
\bibliography{bibliography}
}{
\bibliography{bibliography_short}
}

\ifthenelse{\boolean{long}}{
\newpage
\appendix

\section{Evaluation criteria}\label{app:bewertungskriterien}
\begin{enumerate}
	\item \label{attr:eou} Ease of Use:
	\begin{enumerate}
		\item \label{attr:eou:read} Readability: 
		The length of the function calls or the number of their parameters in a library are crucial for intuitive use \cite{myers2016improving}.
		Function calls with more than three parameters should be well justified and still avoided if possible \cite{martin_clean_2009}.
		The default evaluation determines how many functions have niladic or monadic (less than or equal to two parameters).
		\item \label{attr:eou:defaults} Default Settings: 
		Are there default values for cryptographic procedures and are they secure according to the current status and the recommendations\cite{BSI-TR-2017} of the German Federal Office for Information Security (BSI)?
		\begin{description}
			\item[+2] if the default value corresponds to the specifications.
			\item[-1] if no or bad default values are suggested.
		\end{description}
		\item \label{attr:eou:naming}  Naming Conventions: 
		A naming system is used consistently. 
		For example, for Java libraries the \emph{Google Java Style Guides} \cite{google2015style} or \emph{Oracle Codeconventions}\footnote{\url{https://www.oracle.com/java/technologies/javase/codeconventions-contents.html} Retrieved 16. Feb 2021} are used. 
		The evaluation of this criterion is based on the default evaluation.
		\item \label{attr:eou:reg} Regularity: 
		Attention is paid throughout to the symmetry of names \cite{32713} for the same functionality, pairs such as \emph{connect()} and \emph{disconnect()}.
		\begin{description}
			\item[+2] when naming is used symmetrically where possible.
			\item[ 0] when naming is used symmetrically for central functions.
			\item[-2] when naming does not follow any (recognizable) system.
		\end{description}
		\item \label{attr:eou:selfdesc} Self-describing function names: 
		The function names are understandable, intuitive, and reflect the functionality \cite{32713}. 
		The evaluation of this criterion is based on the default evaluation.
		\end{enumerate}
	\item \label{attr:scale} Scalability:
	\begin{enumerate}
		\item \label{attr:scale:para} Concurrency: 
		Parallel execution of library functions, e.g. in multiple threads or by clusters or load balancers, is generally supported.
Functionalities that are inherently sequential, e.g. encrypting after signing, are excluded from this consideration.
		\begin{description}
			\item[+2] Yes
			\item[ 0] if it is possible via workarounds. For example, by splitting data for encryption into several parts and thus processing them in a distributed manner.
			\item[-2] No
		\end{description}
	\end{enumerate}
	\item \label{attr:test} Testability: 
	\begin{enumerate}
		\item \label{attr:test:test} Testability: 
		There are test recommendations e.g. test classes/functions are offered.
		\begin{description}
			\item[+2] if test functions are supplied, for example, ready-made test classes, default tests, or test examples in the documentation.
			\item[-2] if no test functions or examples are provided in the documentation.
		\end{description}
		\item \label{attr:test:exc} Exceptions:
		Error handling is actively performed by the library.
		\begin{description}
			\item[+2] when customized error handling routines with error descriptions are used.
			\item[ 0] when (standard) error handling routines are processed.
			\item[-2] if there are no error handling routines.
		\end{description}
 	\end{enumerate}
	\item \label{attr:ext} Extendability: 
	\begin{enumerate}
		\item \label{attr:ext:pub} Public: 
		Classes and functions relevant for extended functionality are publicly declared and thus inheritable.
		\begin{description}
			\item[+2] when classes and functions relevant for extended functionality are declared publicly.
			\item[ 0] if partial aspects of the implementation are declared as public.
			\item[-2] if only bundling to no functions and classes are declared public.
		\end{description}
		\item \label{attr:ext:int} Interfaces: 
		Interfaces are used.
		\begin{description}
			\item[+2] for the use of interfaces on all classes that the user is to use.
			\item[ 0] if there are isolated interfaces.
			\item[-2] for the lack of interfaces in the library.
		\end{description}
	\end{enumerate}
	\item \label{attr:fnct} Functional Completeness: 
	\begin{enumerate}
		\item \label{attr:fnct:goal} Purposefulness:
		The library fulfills only its core mission.
		\begin{description}
			\item[+2] when the library fulfills only its core mission.
			\item[ 0] for libraries that provide additional features that are not necessary but useful.
			\item[-2] in libraries where the actual purpose is obscured by non-purposeful features.
		\end{description}
	\end{enumerate}
	\item \label{attr:types} Data Types: 
	\begin{enumerate}
		\item \label{attr:test:ret} Return values: 
		Functions have return values to ensure that they have been executed successfully.
		The evaluation of this criterion is based on the default evaluation.
		\item \label{attr:types:order} Ordering: 
		The order of parameters is consistent.
		The evaluation of this criterion is based on the default evaluation.
	\end{enumerate}
	\item \label{attr:quali} Code Quality:
	Through a SonarCube instance, the code quality is analyzed automatically.
	The index score is composed of U.S. school grades (A through E correspond to 2 through -2) for the following three analyses:
	\begin{enumerate}
	    \item \label{attr:quali:bugs} Bugs
	    \item \label{attr:quali:vulner} Vulnerability
	    \item \label{attr:quali:smell} Code Smell
	\end{enumerate}
	\item \label{attr:cost} Cost: 
	\begin{enumerate}
		\item \label{attr:cost:cost} Cost: 
		The library usable without fees.
		\begin{description}
			\item[+2] for libraries free of cost.
			\item[-2] for libraries with a fee.
		\end{description}
		\item \label{attr:cost:lic} Licence: 
		Under which license is the library offered?
		\begin{description}
			\item[+2] for licenses that allow unrestricted commercial use.
			\item[ 0] for licenses that allow free use for non-commercial purposes or allow commercial use against payment.
			\item[-2] for licenses that do not provide for commercial use and still require payment.
		\end{description}
	\end{enumerate}
	\item \label{attr:req} Requirements:
	\begin{enumerate}
		\item \label{attr:req:pre} Dependencies: 
		There must be other software, software packages, or files installed.
		\begin{description}
			\item[+2] the library automatically installs all dependencies if it has any (possibly via a package manager).
			\item[-2] the dependent software must be installed manually.
		\end{description}
	\end{enumerate}
	\item \label{attr:compl} Complexity:
	\begin{enumerate}
		\item \label{attr:compl:gran} Atomic Setting:
		Precise or fine adjustments can be made.
		\begin{description}
			\item[+2] if an API specifies settings and these can be changed.
			\item[ 0] wenn eine API Einstellungen vorgibt und diese nicht verändert werden können.
			\item[ 0] if an API specifies settings and these cannot be changed.
			\item[-2] when parameters must all be selected manually.
		\end{description}
		\item \label{attr:compl:boil} Boilerplatecode:
		Recurring code overhead must be written by developers in order to use certain functionalities of the API.
		\begin{description}
			\item[+2] for APIs that provide dedicated methods instead of recurring code.
			\item[ 0] for APIs where the largest blocks are simplified.
			\item[-2] when the user has to write the same code over and over again.
		\end{description}
	\end{enumerate}
	\item \label{attr:main} Maintained: 
	\begin{enumerate}
		\item \label{attr:main:freq} Release-frequency:
		How often is a library updated or a new major version released? The question refers to the last three major versions of a library.
		\begin{description}
			\item[+2] there is a fixed release schedule of the library.
			\item[ 0] there is no fixed release schedule, but eventually, the next release will come.
			\item[-2] it is not clear if there will be another release.
		\end{description}
		\item \label{attr:main:patch} Patch frequency:
		Security gaps are quickly fixed by patches.
		\begin{description}
			\item[+2] if patches have been released within 90\footnote{According to the Google "`Project Zero"' deadline. \url{https://en.wikipedia.org/wiki/Project_Zero} Retrieved 28 Oct 2021} days in the past. 
			\item[+1] when patches are delayed for more than 90 days.
			\item[ 0] when it is not clear if patches will be provided.
			\item[-2] if no patches are delivered. 
		\end{description}
		\item \label{attr:main:sup} Support: 
		The library has an official support channel (e.g., a forum, wiki, or mailing lists).
		\begin{description}
			\item[+2] if the project offers free support for the product.
			\item[ 0] with an official paid support.
			\item[-2] when there are no official information channels and sources.
		\end{description}
	\end{enumerate}
	\item \label{attr:spr} Spread:
	\begin{enumerate}
		\item \label{attr:spr:succ} Successful Stories:
		There are articles on successful stories of using the library in other projects. 
		Successful stories are articles, contributions or scientific papers on successful use, which, for example, attest to the added value of the library or describe experiences.
		\begin{description}
			\item[+2] in the case of several articles or field reports in reputable trade journals such as heise.de\footnote{\url{https://www.heise.de/} Retrieved 28 Oct 2021}.
			\item[ 0] with no significant mentions in professional journals.
			\item[-2] for purchased posts on blogs or similar formats.
		\end{description}
        \item \label{attr:spr:git} Repositories: 
				The project is very popular on repositories (such as GitHub, GitLab, SourceForge, Bitbucket or a repository managed by the project).
				This can be read off from "`likes"', "`stars"' or similar. 
				The basis for the rating is the most popular Java library on GitHub in English language at the time of the work (December 2020) (Mindustry with 7557 stars\footnote{\url{https://github.com/trending/java?since=monthly&spoken_language_code=en} Retrieved 23 Dec 2020}).
		The evaluation of this criterion is based on the default evaluation.

	\end{enumerate}
	\item \label{attr:perf} Performance Impact

	\item \label{attr:sec} Security
	\begin{enumerate}
		\item \label{attr:sec:std} Standards: 
		Are used exclusively algorithms that are on the white list \cite{BSI-TR-2017}.
		\begin{description}
			\item[+2] when algorithms and standards meet (or exceed) the current state of the art.
			\item[ 0] if the library meets the standards exactly.
			\item[-2] if a sub-aspect is obsolete or is not considered safe enough.
		\end{description}
		\item \label{attr:sec:cert} Certificated: 
		Has the library been certified?
		\begin{description}
	    	\item[+2] Yes, multiple.
			\item[ 0] Yes.
			\item[-2] No.
		\end{description}
	\end{enumerate}
	\item \label{attr:doc} Documentation:
	\begin{enumerate}
		\item \label{attr:doc:fkt} Function documentation: 
		Each method or function is documented. 
 
		The evaluation of this criterion is based on the default evaluation.
		\item \label{attr:doc:ex} Examples: 
		There is an example of the correct application for each method or function.
		The evaluation of this criterion is based on the default evaluation.
	\end{enumerate}
\end{enumerate}

\input{interview-qst.tex}
\input{survey.tex}

\small
\input{metriken-small.tex}
\normalsize 
}{
}
\end{document}

%% file: interview-qst.tex
\section{Interview}\label{inter:qst}
This is the schedule of interviews with decision makers presented in Section~\ref{ch:attribute:interview}.
\begin{enumerate}
    \item \label{inter:qst:1} 
		What is your position or role in projects?
    \item \label{inter:qst:2} 
		How long have you had experience as in this position or role?
    \item \label{inter:qst:3} 
		Have you ever worked with a crypto API?
    \begin{enumerate}
        \item 
				From now on, assume you need to choose a Crypto API for your next project.
    \end{enumerate}
    \item \label{inter:qst:4} 
		Do you have a decision process for selecting a (crypto) API in your projects?
    \begin{enumerate}
        \item 
				What does this process look like?
    \end{enumerate}
    \item \label{inter:qst:5} 
		How do you plan to proceed?
    \item \label{inter:qst:6} 
		Do you have to adhere to specifications?
    \item \label{inter:qst:7} 
		Who is involved in this procedure? (How many?)
    \begin{enumerate}
        \item 
				How are the roles distributed in the team?
    \end{enumerate}
    \item \label{inter:qst:8} 
		Do you get help from outside the team, consultants, or external collaborators?
    \begin{enumerate}
        \item 
				What information do you get from them?
    \end{enumerate}
    \item \label{inter:qst:9} 
		What are your steps for making the decision?
    \item \label{inter:qst:10} 
		Where do you get the information from?
    \item \label{inter:qst:11} 
		What criteria do you use?
    \begin{enumerate}
        \item 
				Where do they come from?
        \item 
				Why do you attach importance to these topics?
        \item 
				How do you evaluate the criteria?
        \item 
				How do you weight the criteria?
        \begin{enumerate}
            \item 
						Are the weightings regularly reviewed and adjusted?
        \end{enumerate}
        \item 
				Are the criteria given to you?
    \end{enumerate}
    \item \label{inter:qst:12} 
		Do you have to justify your decision?
    \item \label{inter:qst:13} 
		Do you have any questions yourself or would you like to share something?
\end{enumerate}

%% file: survey.tex
\section{Questionnaire}\label{appendix:fragebogen}
The questionnaire from the online survey presented in Section~\ref{ch:attribute:validierung}.
\begin{enumerate}
	\item \label{qst:qst:1} 
	For how long have you worked as a developer in years? (number)
	\item \label{qst:qst:2} 
	How often do you need to choose an API? (checkbox: never, once a year, several times a year, once a month, several times a month, weekly)
	\item \label{qst:qst:3} 
	Have you used a crypto-API before? (yes/no)
	\item \label{qst:qst:4} 
	Does the description match the term? (Answer options: not at all, barely, mediocrally, fairly, exceptionally)
	List with the attributes and descriptions from table \ref{table:score:atts}.
	\item \label{qst:qst:5} 
	Is the term suitable for the rating of an API (not in context of own projects)? (Answer options: not at all, barely, mediocrally, fairly, exceptionally)
	List with attributes from table \ref{table:score:atts}.
	\item \label{qst:qst:6} 
	Sorting question: Sort the following attributes according to their relevance for choosing a suitable (crypto-)API for your last project in descending order. 
	Drack-and-drop list with the attributes from table \ref{table:score:atts}. 
	\item \label{qst:qst:8} 
	Do you have any Remarks, Feedback or suggestions concerning the questions or survey?
\end{enumerate}

%% file: metriken-small.tex
\clearpage
\onecolumn
\section{Structure of attributes}
\begin{longtable}{|llllp{6cm}|}
	\caption{Hierarchical structure of attributes obtained from literature and interviews. The notation ``I'' represents the source of an interview with the number of the mention in parentheses. Otherwise, the source is given from the literature. }\label{tab:allAttributes} \\

\hline
    ID    & Level 1 & Level 2 & Level 3 & Level 4 \\
\hline
\endfirsthead 
\caption{Continued: Hierarchical structure of attributes obtained from the literature and interviews.}\\
\hline
    ID    & Level 1 & Level 2 & Level 3 & Level 4 \\
\hline
\endhead 
\multicolumn{5}{r}{Continue on next page} \\
\endfoot
\hline
\endlastfoot

	\hline	
    1     & \multicolumn{4}{l|}{Ease of Use (\cite{7965488}, I:3)} \\
    \hline
    1.1   &       & \multicolumn{3}{l|}{Flexibility and efficiency of Use (\cite{myers2016improving}, I:1)}  \\
    \hline
    1.2   &       & \multicolumn{3}{l|}{Integrierbarkeit (\cite{32713}, I:3)}  \\
    \hline
    1.2.1 &       &       & \multicolumn{2}{l|}{Role Expressiveness (\cite{clarke_measuring_nodate})}  \\
    \hline
    1.3   &       & \multicolumn{3}{l|}{Reusability (\cite{32713}) }  \\
    \hline
    1.4   &       & \multicolumn{3}{l|}{Combinability (I:1)}  \\
    \hline
    1.5   &       & \multicolumn{3}{l|}{Domain Correspondence (\cite{clarke_measuring_nodate})}    \\
    \hline
    1.6   &       & \multicolumn{3}{l|}{Compatibility (\cite{7965488})}   \\
    \hline
    1.6.1 &       &       & \multicolumn{2}{l|}{Substitutibility (\cite{32713})}  \\
    \hline
    1.7   &       & \multicolumn{3}{l|}{Guidance (\cite{acar2017comparing})} \\
    \hline
    1.7.1 &       &       & \multicolumn{2}{l|}{Conceptual Correctness (\cite{grill2012methods})}  \\
    \hline
    1.7.1.1 &       &       &       & Self-Describing Functions (\cite{32713})   \\
    \hline
    1.7.1.2 &       &       &       & Recognition (\cite{myers2016improving})   \\
    \hline
    1.7.1.3 &       &       &       & Naming Consistency (\cite{32713}, I:2)   \\
    \hline
    1.7.1.4 &       &       &       & Match between system and real world (\cite{myers2016improving})   \\
    \hline
    1.7.2 &       &       & \multicolumn{2}{l|}{Error Prevention (\cite{myers2016improving})} \\
    \hline
    1.7.3 &       &       & \multicolumn{2}{l|}{Predictable (\cite{32713})}  \\
    \hline
    1.7.3.1 &       &       &       & Safe and Secure Defaults (\cite{acar2017comparing})   \\
    \hline
    1.7.4 &       &       & \multicolumn{2}{l|}{Caller's Perspective (\cite{grill2012methods})}  \\
    \hline
    1.8   &       & \multicolumn{3}{l|}{Readability (\cite{32713})} \\
    \hline
    1.8.1 &       &       & \multicolumn{2}{l|}{Naming (\cite{grill2012methods})}  \\
    \hline
    1.9   &       & \multicolumn{3}{l|}{Learning curve (I:2) }  \\
    \hline
    1.10  &       & \multicolumn{3}{l|}{Penetrability (\cite{clarke_measuring_nodate})} \\
    \hline
    1.11  &       & \multicolumn{3}{l|}{Learning Style (\cite{clarke_measuring_nodate})}  \\
    \hline
    1.12  &       & \multicolumn{3}{l|}{Work-Step Unit (\cite{clarke_measuring_nodate}) } \\
    \hline
    1.12.1 &       &       & \multicolumn{2}{l|}{Leftovers for Client Code (\cite{grill2012methods})}  \\
    \hline
    1.12.2 &       &       & \multicolumn{2}{l|}{Short Chain of References (\cite{grill2012methods})}  \\
    \hline
    1.13  &       & \multicolumn{3}{l|}{Premature Commitment (\cite{clarke_measuring_nodate})} \\
    \hline
    1.13.1 &       &       & \multicolumn{2}{l|}{API Elaboration (\cite{clarke_measuring_nodate})}  \\
    \hline
    1.14  &       & \multicolumn{3}{p{9cm}|}{Consistency, standards and Conventions (\cite{myers2016improving,grill2012methods,clarke_measuring_nodate,32713,ko2004six})} \\
    \hline
    1.15  &       & \multicolumn{3}{l|}{User Control and Freedom (\cite{myers2016improving})}  \\
    \hline
	\hline
    2     & \multicolumn{4}{l|}{Scalability (I:2)}   \\
    \hline
    2.1   &       & \multicolumn{3}{l|}{Synchrony (\cite{7965488})} \\
    \hline
    2.2   &       & \multicolumn{3}{l|}{Handleable Datasize (I:2) }  \\
    \hline
    2.3   &       & \multicolumn{3}{l|}{Concurrency (\cite{grill2012methods})}  \\
    \hline
	\hline
    3     & \multicolumn{4}{l|}{Testability (\cite{7965488}, I:1)}   \\
    \hline
    3.1   &       & \multicolumn{3}{p{9cm}|}{Error handling and visibility of system status (\cite{grill2012methods,ko2004six,myers2016improving})}  \\
    \hline
    3.1.1 &       &       & \multicolumn{2}{l|}{Error Checking and Responsiveness (\cite{7965488})}  \\
    \hline
    3.2   &       & \multicolumn{3}{l|}{Progressive evaluation (\cite{clarke_measuring_nodate})}  \\
    \hline
    3.3   &       & \multicolumn{3}{l|}{Error-Reporting (\cite{32713})}  \\
    \hline
    3.3.1 &       &       & \multicolumn{2}{l|}{Exception Indication (\cite{32713})}  \\
    \hline
    3.3.2 &       &       & \multicolumn{2}{l|}{Helpful error messaging and logging (\cite{myers2016improving})}  \\
    \hline
    3.4   &       & \multicolumn{3}{l|}{Debbuging (\cite{ko2004six}, I:1)} \\
    \hline
	\hline
    4     & \multicolumn{4}{l|}{Extendibility (\cite{32713}, I:1)} \\
    \hline
    4.1   &       & \multicolumn{3}{l|}{API Evolvability (\cite{7965488})}  \\
    \hline
    4.2   &       & \multicolumn{3}{l|}{API Viscosity (\cite{clarke_measuring_nodate})}  \\
    \hline
	\hline
    5     & \multicolumn{4}{l|}{Functional Completeness (\cite{7965488})}   \\
    \hline
    5.1   &       & \multicolumn{3}{l|}{Features (\cite{acar2017comparing})}  \\
    \hline
    5.2   &       & \multicolumn{3}{l|}{Purposefulness/Light Footprint (\cite{32713,32713})}  \\
    \hline
    5.3   &       & \multicolumn{3}{l|}{Functionality (\cite{32713})} \\
    \hline
	\hline
    6     & \multicolumn{4}{l|}{Data Types (\cite{grill2012methods})}  \\
    \hline
    6.1   &       & \multicolumn{3}{l|}{Method Parameters and Return Types (\cite{32713,grill2012methods,ko2004six})}  \\
    \hline
    6.2   &       & \multicolumn{3}{l|}{Consistent Parameter Ordering (I:1) } \\
    \hline
	\hline
    7     & \multicolumn{4}{l|}{Code Quality (I:1)}   \\
    \hline
    7.1   &       & \multicolumn{3}{l|}{Standards (I:4)}  \\
    \hline
    7.2   &       & \multicolumn{3}{l|}{Patterns/interfaces (\cite{grill2012methods})}  \\
    \hline
	\hline
    8     & \multicolumn{4}{l|}{Cost (I:7)}   \\
    \hline
    8.1   &       & \multicolumn{3}{l|}{Copyright/License (I:4)} \\
    \hline
	\hline
    9     & \multicolumn{4}{l|}{Requirements (I:3)}  \\
    \hline
    9.1   &       & \multicolumn{3}{l|}{Dependency (I:2)} \\
    \hline
	\hline
    10    & \multicolumn{4}{l|}{Complexity (\cite{7965488,grill2012methods}, I:2)}  \\
    \hline
    10.1  &       & \multicolumn{3}{l|}{Aesthetic and minimalist design (\cite{myers2016improving})}  \\
    \hline
    10.1.1 &       &       & \multicolumn{2}{l|}{Information Hiding (\cite{32713})}  \\
    \hline
    10.2  &       & \multicolumn{3}{l|}{Working Framework (\cite{clarke_measuring_nodate})}\\
    \hline
    10.3  &       & \multicolumn{3}{l|}{Abstraction Level (\cite{clarke_measuring_nodate})}  \\
    \hline
    10.4  &       & \multicolumn{3}{l|}{Atomic Setting (\cite{7965488})}\\
    \hline
    10.5  &       & \multicolumn{3}{l|}{Boilerplate Code (\cite{32713})}\\
    \hline
    10.6  &       & \multicolumn{3}{l|}{Simplicity (\cite{acar2017comparing,32713}, I:1)}  \\
    \hline
    10.6.1 &       &       & \multicolumn{2}{l|}{Single way to do one thing (\cite{grill2012methods})}  \\
    \hline
	\hline
    11    & \multicolumn{4}{l|}{Maintained (I:6)}  \\
    \hline
    11.1  &       & \multicolumn{3}{l|}{Support (I:2)} \\
    \hline
	\hline
    12    & \multicolumn{4}{l|}{Spread (I:16)} \\
    \hline
    12.1  &       & \multicolumn{3}{l|}{Community (I:2)}  \\
    \hline
    12.1.1 &       &       & \multicolumn{2}{l|}{Stackoverflow (I:2)}  \\
    \hline
    12.2  &       & \multicolumn{3}{l|}{Reputation(I:3)}  \\
    \hline
    12.3  &       & \multicolumn{3}{l|}{Recommendations(I:1)} \\
    \hline
    12.3.1 &       &       & \multicolumn{2}{l|}{Success Stories(I:4)}  \\
    \hline
	\hline
    13    & \multicolumn{4}{l|}{Performance Impact (\cite{32713})}  \\
    \hline
    13.1  &       & \multicolumn{3}{l|}{Latency (\cite{7965488})} \\
    \hline
	\hline
    14    & \multicolumn{4}{l|}{Security (\cite{7965488})}  \\
    \hline
    14.1  &       & \multicolumn{3}{l|}{Security (I:1)} \\
    \hline
    14.2  &       & \multicolumn{3}{l|}{Data Governance (I:1)} \\
    \hline
    14.3  &       & \multicolumn{3}{l|}{Used process (I:2)} \\
    \hline
	\hline
    15    & \multicolumn{4}{l|}{Documentation (\cite{grill2012methods,myers2016improving,ko2004six,32713,acar2017comparing}, I:14)}  \\
    \hline
    15.1  &       & \multicolumn{3}{l|}{Best Practices (I:1)}  \\
    \hline
    15.2  &       & \multicolumn{3}{l|}{Examples (I:4)}  \\
    \hline
  
\end{longtable}

\clearpage
\twocolumn

%% file: main.bbl

\begin{thebibliography}{20}


\ifx \showCODEN    \undefined \def \showCODEN     #1{\unskip}     \fi
\ifx \showDOI      \undefined \def \showDOI       #1{#1}\fi
\ifx \showISBNx    \undefined \def \showISBNx     #1{\unskip}     \fi
\ifx \showISBNxiii \undefined \def \showISBNxiii  #1{\unskip}     \fi
\ifx \showISSN     \undefined \def \showISSN      #1{\unskip}     \fi
\ifx \showLCCN     \undefined \def \showLCCN      #1{\unskip}     \fi
\ifx \shownote     \undefined \def \shownote      #1{#1}          \fi
\ifx \showarticletitle \undefined \def \showarticletitle #1{#1}   \fi
\ifx \showURL      \undefined \def \showURL       {\relax}        \fi
\providecommand\bibfield[2]{#2}
\providecommand\bibinfo[2]{#2}
\providecommand\natexlab[1]{#1}
\providecommand\showeprint[2][]{arXiv:#2}

\bibitem[\protect\citeauthoryear{Acar, Backes, Fahl, Garfinkel, Kim, Mazurek,
  and Stransky}{Acar et~al\mbox{.}}{2017}]%
        {acar2017comparing}
\bibfield{author}{\bibinfo{person}{Yasemin Acar}, \bibinfo{person}{Michael
  Backes}, \bibinfo{person}{Sascha Fahl}, \bibinfo{person}{Simson Garfinkel},
  \bibinfo{person}{Doowon Kim}, \bibinfo{person}{Michelle~L. Mazurek}, {and}
  \bibinfo{person}{Christian Stransky}.} \bibinfo{year}{2017}\natexlab{}.
\newblock \showarticletitle{Comparing the Usability of Cryptographic APIs}. In
  \bibinfo{booktitle}{\emph{2017 IEEE Symposium on Security and Privacy (SP)}}.
  \bibinfo{publisher}{IEEE}, \bibinfo{address}{New York, USA},
  \bibinfo{pages}{154--171}.
\newblock
\urldef\tempurl%
\url{https://doi.org/10.1109/SP.2017.52}
\showDOI{\tempurl}


\bibitem[\protect\citeauthoryear{Bloch}{Bloch}{2006}]%
        {32713}
\bibfield{author}{\bibinfo{person}{Joshua Bloch}.}
  \bibinfo{year}{2006}\natexlab{}.
\newblock \showarticletitle{How to design a good API and why it matters}. In
  \bibinfo{booktitle}{\emph{Proc. 21st ACM SIGPLAN Conference (OOPSLA)}}.
  \bibinfo{publisher}{ACM}, \bibinfo{address}{Portland, Oregon},
  \bibinfo{pages}{506--507}.
\newblock
\urldef\tempurl%
\url{https://doi.org/10.1145/1176617.1176622}
\showDOI{\tempurl}


\bibitem[\protect\citeauthoryear{BSI}{BSI}{2021}]%
        {BSI-TR-2017}
\bibfield{author}{\bibinfo{person}{BSI}.} \bibinfo{year}{2021}\natexlab{}.
\newblock \bibinfo{booktitle}{\emph{Kryptographische Verfahren: Empfehlungen
  und Schlüssellängen}}.
\newblock \bibinfo{type}{{T}echnical {R}eport} BSI TR-02102-1.
  \bibinfo{institution}{BSI}, \bibinfo{address}{Bonn, Germany}.
\newblock
\urldef\tempurl%
\url{https://www.bsi.bund.de/SharedDocs/Downloads/DE/BSI/Publikationen/TechnischeRichtlinien/TR02102/BSI-TR-02102.pdf}
\showURL{%
\tempurl}


\bibitem[\protect\citeauthoryear{Clarke}{Clarke}{2004}]%
        {clarke_measuring_nodate}
\bibfield{author}{\bibinfo{person}{Steven Clarke}.}
  \bibinfo{year}{2004}\natexlab{}.
\newblock \bibinfo{title}{Measuring {API} {Usability}}.
\newblock
\newblock
\urldef\tempurl%
\url{http://www.drdobbs.com/windows/measuring-api-usability/184405654}
\showURL{%
\tempurl}


\bibitem[\protect\citeauthoryear{Gao, Chen, Wu, and Gao}{Gao
  et~al\mbox{.}}{2015}]%
        {gao_manifold-learning_2015}
\bibfield{author}{\bibinfo{person}{Wei Gao}, \bibinfo{person}{Liang Chen},
  \bibinfo{person}{Jian Wu}, {and} \bibinfo{person}{Honghao Gao}.}
  \bibinfo{year}{2015}\natexlab{}.
\newblock \showarticletitle{Manifold-{Learning} {Based} {API} {Recommendation}
  for {Mashup} {Creation}}. In \bibinfo{booktitle}{\emph{2015 {IEEE}
  {International} {Conference} on {Web} {Services}}}.
  \bibinfo{publisher}{IEEE}, \bibinfo{address}{New York, USA},
  \bibinfo{pages}{432--439}.
\newblock
\urldef\tempurl%
\url{https://doi.org/10.1109/ICWS.2015.64}
\showDOI{\tempurl}


\bibitem[\protect\citeauthoryear{google}{google}{2015}]%
        {google2015style}
\bibfield{author}{\bibinfo{person}{google}.} \bibinfo{year}{2015}\natexlab{}.
\newblock \bibinfo{booktitle}{\emph{Google Java Style Guide}}.
\newblock Google.
\newblock
\urldef\tempurl%
\url{https://google.github.io/styleguide/javaguide.html}
\showURL{%
\tempurl}


\bibitem[\protect\citeauthoryear{Grill, Polacek, and Tscheligi}{Grill
  et~al\mbox{.}}{2012}]%
        {grill2012methods}
\bibfield{author}{\bibinfo{person}{Thomas Grill}, \bibinfo{person}{Ondrej
  Polacek}, {and} \bibinfo{person}{Manfred Tscheligi}.}
  \bibinfo{year}{2012}\natexlab{}.
\newblock \showarticletitle{Methods towards API usability: a structural
  analysis of usability problem categories}. In
  \bibinfo{booktitle}{\emph{International conference on human-centred software
  engineering}}. \bibinfo{publisher}{Springer}, \bibinfo{address}{Berlin},
  \bibinfo{pages}{164--180}.
\newblock


\bibitem[\protect\citeauthoryear{Hedderich and Sachs}{Hedderich and
  Sachs}{2016}]%
        {hedderich2016angewandte}
\bibfield{author}{\bibinfo{person}{Jürgen Hedderich} {and}
  \bibinfo{person}{Lothar Sachs}.} \bibinfo{year}{2016}\natexlab{}.
\newblock \bibinfo{booktitle}{\emph{Angewandte Statistik}}.
\newblock \bibinfo{publisher}{Springer}, \bibinfo{address}{Berlin}.
\newblock


\bibitem[\protect\citeauthoryear{KG}{KG}{2015}]%
        {BSI.2015}
\bibfield{author}{\bibinfo{person}{Rohde \& Schwarz GmbH \&~Co. KG}.}
  \bibinfo{year}{2015}\natexlab{}.
\newblock \bibinfo{title}{Sichere Implementierung einer allgemeinen
  Kryptobibliothek: Arbeitspaket 1: Sichtung und Analyse bestehender
  Kryptobibliotheken}.
\newblock
\newblock
\urldef\tempurl%
\url{https://media.frag-den-staat.de/files/foi/89304/Analyse_geschNANAMEErzt_Vorblatt.pdf}
\showURL{%
\tempurl}


\bibitem[\protect\citeauthoryear{Ko, Myers, and Aung}{Ko et~al\mbox{.}}{2004}]%
        {ko2004six}
\bibfield{author}{\bibinfo{person}{Andrew~J Ko}, \bibinfo{person}{Brad~A
  Myers}, {and} \bibinfo{person}{Htet~Htet Aung}.}
  \bibinfo{year}{2004}\natexlab{}.
\newblock \showarticletitle{Six learning barriers in end-user programming
  systems}. In \bibinfo{booktitle}{\emph{2004 IEEE Symposium on Visual
  Languages-Human Centric Computing}}. \bibinfo{publisher}{IEEE},
  \bibinfo{address}{New York, USA}, \bibinfo{pages}{199--206}.
\newblock


\bibitem[\protect\citeauthoryear{Kuckartz}{Kuckartz}{2018}]%
        {kuckartz_qualitative_2018}
\bibfield{author}{\bibinfo{person}{Udo Kuckartz}.}
  \bibinfo{year}{2018}\natexlab{}.
\newblock \bibinfo{booktitle}{\emph{Qualitative {Inhaltsanalyse}: {Methoden},
  {Praxis}, {Computerunterstützung}} (\bibinfo{edition}{4. auflage} ed.)}.
\newblock \bibinfo{publisher}{Beltz Juventa}, \bibinfo{address}{Weinheim
  Basel}.
\newblock
\showISBNx{978-3-7799-3682-4 978-3-7799-4683-0}


\bibitem[\protect\citeauthoryear{Lazar, Feng, and Hochheiser}{Lazar
  et~al\mbox{.}}{2017}]%
        {lazar_research_2017}
\bibfield{author}{\bibinfo{person}{Jonathan Lazar},
  \bibinfo{person}{Jinjuan~Heidi Feng}, {and} \bibinfo{person}{Harry
  Hochheiser}.} \bibinfo{year}{2017}\natexlab{}.
\newblock \bibinfo{booktitle}{\emph{Research {Methods} in {Human}-{Computer}
  {Interaction}}}.
\newblock \bibinfo{publisher}{Morgan Kaufmann}, \bibinfo{address}{Burlington,
  Massachusetts}.
\newblock
\showISBNx{978-0-12-809343-6}
\newblock
\shownote{Google-Books-ID: hbkxDQAAQBAJ.}


\bibitem[\protect\citeauthoryear{Martin}{Martin}{2009}]%
        {martin_clean_2009}
\bibfield{author}{\bibinfo{person}{R.C. Martin}.}
  \bibinfo{year}{2009}\natexlab{}.
\newblock \bibinfo{booktitle}{\emph{Clean {Code}: {A} {Handbook} of {Agile}
  {Software} {Craftsmanship}}}.
\newblock \bibinfo{publisher}{Prentice Hall}, \bibinfo{address}{Upper Saddle
  River, New Jersey}.
\newblock
\showISBNx{978-0-13-235088-4}
\urldef\tempurl%
\url{https://books.google.de/books?id=dwSfGQAACAAJ}
\showURL{%
\tempurl}


\bibitem[\protect\citeauthoryear{Mayring}{Mayring}{2000}]%
        {mayring_qualitative_2000}
\bibfield{author}{\bibinfo{person}{Philipp Mayring}.}
  \bibinfo{year}{2000}\natexlab{}.
\newblock \showarticletitle{Qualitative {Inhaltsanalyse}}. In
  \bibinfo{booktitle}{\emph{Forum {Qualitative} {Sozialforschung}/{Forum}:
  {Qualitative} {Social} {Research}}}, Vol.~\bibinfo{volume}{1}.
  \bibinfo{publisher}{Institut für Qualitative Forschung},
  \bibinfo{address}{Berlin}.
\newblock


\bibitem[\protect\citeauthoryear{Myers and Stylos}{Myers and Stylos}{2016}]%
        {myers2016improving}
\bibfield{author}{\bibinfo{person}{Brad~A Myers} {and} \bibinfo{person}{Jeffrey
  Stylos}.} \bibinfo{year}{2016}\natexlab{}.
\newblock \showarticletitle{Improving API usability}.
\newblock \bibinfo{journal}{\emph{Commun. ACM}} \bibinfo{volume}{59},
  \bibinfo{number}{6} (\bibinfo{year}{2016}), \bibinfo{pages}{62--69}.
\newblock


\bibitem[\protect\citeauthoryear{Rohrmann}{Rohrmann}{1978}]%
        {rohr1978skalen}
\bibfield{author}{\bibinfo{person}{B. Rohrmann}.}
  \bibinfo{year}{1978}\natexlab{}.
\newblock \showarticletitle{Empirische Studien zur Entwicklung von
  Antwortskalen für die sozialwissenschaftliche Forschung}.
\newblock \bibinfo{journal}{\emph{Zs. für Sozialpsychologie}}
  \bibinfo{volume}{9} (\bibinfo{year}{1978}), \bibinfo{pages}{222--245}.
\newblock


\bibitem[\protect\citeauthoryear{Scheller and Kuehn}{Scheller and
  Kuehn}{2015}]%
        {SchellerKuehnAutomated}
\bibfield{author}{\bibinfo{person}{Thomas Scheller} {and} \bibinfo{person}{Eva
  Kuehn}.} \bibinfo{year}{2015}\natexlab{}.
\newblock \showarticletitle{Automated measurement of API usability: The API
  Concepts Framework}.
\newblock \bibinfo{journal}{\emph{Information and Software Technology}}
  \bibinfo{volume}{61} (\bibinfo{date}{02} \bibinfo{year}{2015}).
\newblock
\urldef\tempurl%
\url{https://doi.org/10.1016/j.infsof.2015.01.009}
\showDOI{\tempurl}


\bibitem[\protect\citeauthoryear{Wilde and Amundsen}{Wilde and
  Amundsen}{2019}]%
        {wilde_challenge_2019}
\bibfield{author}{\bibinfo{person}{Erik Wilde} {and} \bibinfo{person}{Mike
  Amundsen}.} \bibinfo{year}{2019}\natexlab{}.
\newblock \showarticletitle{The {Challenge} of {API} {Management}: {API}
  {Strategies} for {Decentralized} {API} {Landscapes}}. In
  \bibinfo{booktitle}{\emph{Companion {Proceedings} of {The} 2019 {World}
  {Wide} {Web} {Conference}}}. \bibinfo{publisher}{ACM}, \bibinfo{address}{San
  Francisco, USA}, \bibinfo{pages}{1327--1328}.
\newblock
\urldef\tempurl%
\url{https://doi.org/10.1145/3308560.3320089}
\showDOI{\tempurl}


\bibitem[\protect\citeauthoryear{Xie, Liu, Tang, Zhou, Cao, and Shi}{Xie
  et~al\mbox{.}}{2016}]%
        {xie_multi-relation_2016}
\bibfield{author}{\bibinfo{person}{Fenfang Xie}, \bibinfo{person}{Jianxun Liu},
  \bibinfo{person}{Mingdong Tang}, \bibinfo{person}{Dong Zhou},
  \bibinfo{person}{Buqing Cao}, {and} \bibinfo{person}{Min Shi}.}
  \bibinfo{year}{2016}\natexlab{}.
\newblock \showarticletitle{Multi-relation {Based} {Manifold} {Ranking}
  {Algorithm} for {API} {Recommendation}}. In
  \bibinfo{booktitle}{\emph{Advances in {Services} {Computing}}}.
  \bibinfo{publisher}{Springer}, \bibinfo{address}{Cham},
  \bibinfo{pages}{15--32}.
\newblock
\urldef\tempurl%
\url{https://doi.org/10.1007/978-3-319-49178-3_2}
\showDOI{\tempurl}


\bibitem[\protect\citeauthoryear{{Zghidi}, {Hammouda}, {Hnich}, and
  {Knauss}}{{Zghidi} et~al\mbox{.}}{2017}]%
        {7965488}
\bibfield{author}{\bibinfo{person}{A. {Zghidi}}, \bibinfo{person}{I.
  {Hammouda}}, \bibinfo{person}{B. {Hnich}}, {and} \bibinfo{person}{E.
  {Knauss}}.} \bibinfo{year}{2017}\natexlab{}.
\newblock \showarticletitle{On the Role of Fitness Dimensions in API Design
  Assessment - An Empirical Investigation}. In \bibinfo{booktitle}{\emph{2017
  IEEE/ACM 1st International Workshop on API Usage and Evolution (WAPI)}}.
  \bibinfo{publisher}{IEEE}, \bibinfo{address}{New York, USA},
  \bibinfo{pages}{19--22}.
\newblock


\end{thebibliography}
